# Single-shot measurement of few-cycle optical waveforms on a chip

Yangyang Liu[1], John E. Beetar[1], Jonathan Nesper[1], Shima Gholam-Mirzaei[1,2], and Michael Chini[1,3*]

[1]*Department of Physics, University of Central Florida, Orlando FL 32816, USA*

[2]*Joint Attosecond Science Laboratory (JASLab), National Research Council of Canada and University of Ottawa, Ottawa ON K1A0R6, Canada*

[3]*CREOL, the College of Optics and Photonics, University of Central Florida, Orlando FL 32816, USA*

*\*E-mail: Michael.Chini@ucf.edu*

**The measurement of transient optical fields has proven critical to understanding the dynamical mechanisms underlying ultrafast physical and chemical phenomena, and is key to realizing higher speeds in electronics and telecommunications. Complete characterization of optical waveforms, however, requires an 'optical oscilloscope' capable of resolving the electric field oscillations with sub-femtosecond resolution and with single-shot operation. Here, we show that strong-field nonlinear excitation of photocurrents in a silicon-based image sensor chip can provide the sub-cycle optical gate necessary to characterize carrier-envelope phase-stable optical waveforms in the mid-infrared. By mapping the temporal delay between an intense excitation and weak perturbing pulse onto a transverse spatial coordinate of the image sensor, we show that the technique allows single-shot measurement of few-cycle waveforms.**

The development, characterization, and control of intense, few-cycle optical fields have enabled the generation of isolated attosecond pulses [1] and ushered in a new regime of field-resolved optical spectroscopy and microscopy (or "fieldoscopy" [2]) techniques [3-5]. These advances, in turn, have stimulated the development of pulse characterization tools capable of measuring the complex electric field – including the carrier-envelope (or "absolute") phase [6] – of a few-cycle laser field. To date, these techniques have been based either on field-driven "streaking" of photoionized electrons [7,8], electro-optic sampling (EOS) [9,10], or the optical perturbation of a strong-field response [11], and in all cases have required measurements over many laser shots in a scanning geometry. The scanning geometry, however, requires that all pulses emitted from the laser are essentially identical [12] and that the repetition rate of the laser is sufficiently high to allow for a multi-shot measurement.

The primary requirement for complete characterization of an optical waveform is the generation of an ultrafast gate, with a duration below the half-cycle duration of the field oscillation to be measured. In most cases, including EOS and attosecond streaking, this has been achieved using few-cycle sampling pulses with much shorter wavelengths than the fields to be characterized. However, it has also been shown that a strong-field response, such as tunnel ionization [11], high-order harmonic generation [13], or multiphoton excitation in solids [14], can provide a sub-cycle gate. With this, in analogy to pulse measurement schemes based on perturbative nonlinear optics [15,16], it becomes possible to use the laser field itself as the measurement tool. Here, we show that nonlinear photocurrents, generated in a silicon-based image sensor by intense mid-infrared pulses, can provide such an ultrafast gate. By mapping the time-delay onto the transverse position [17], we further take advantage of the spatial resolution inherent to image sensors to realize *single-shot* detection of mid-infrared laser waveforms.

The experimental measurement, which is an extension of the TIPTOE (tunneling ionization with a perturbation for the time-domain observation of an electric field) technique [11, 14], is described schematically in Fig. 1a and in the Methods. Briefly, an intense "fundamental" pulse with a central wavelength of 3.4 µm creates charge packets in the pixels of a silicon-based image sensor via multiphoton excitation, leading to detectable photocurrents. The probability of excitation is perturbed by the field of a weak "perturbation" pulse, leading to a modulation in the excitation probability and therefore in the magnitude of the detected photocurrent. Previously we have shown that for collinear fundamental and perturbation pulses, the dependence of the modulation in the excitation probability on the time delay between the two pulses encodes the time-varying electric field waveform of the laser pulse [14]. Here, by using a crossed-beam geometry with cylindrical focusing, we map the time delay onto a transverse spatial coordinate of the image sensor chip to achieve single-shot detection.

The experimental image obtained from the sensor chip is shown in Fig. 1b, along with lineouts showing the magnitude of the photocurrent as a function of the spatial position on the detector with and without the perturbation pulse present. Here, the image and lineouts represent a true single-shot measurement, without integrating over multiple pulses or averaging multiple images. The signal is dominated by multiphoton excitation by the fundamental pulse, and the perturbation pulse produces no photocurrent on its own. Instead, the presence of the perturbation pulse is observed as a weak modulation on the signal produced by the fundamental pulse. Importantly, this modulation does not arise due to linear interference between the fundamental and perturbation pulses, but rather as a cross-correlation between the perturbation pulse and a sub-cycle electro-optic gate arising due to the multiphoton excitation of charge packets in the image sensor chip [18]. From the perturbed photocurrent signal, the modulation waveform of the perturbation pulse (Fig. 1c) can be obtained by subtracting the fundamental-only signal, followed by a normalization process to remove the effects of the spatial variation of the fundamental beam profile (see Methods and Supplementary Information), and higher signal-to-noise ratio can be obtained by averaging over multiple identical laser shots. Finally, we take the Fourier transform of the modulation waveform to obtain the frequency spectrum, which can be compared to an independent measurement of the spectrum, and the spectral phase.

Our previous work using band fluorescence to detect the modulation of the multiphoton excitation probability showed that the modulated signals in Fig. 1 reflect the electric field envelope and the time-dependent phase (i.e. the carrier frequency and the frequency chirp) of the perturbation pulse, even for multi-cycle pulses. That is to say, the modulation provides a *pulse* measurement but not necessarily a *waveform* measurement. In those measurements, as was the case in prior work using nonlinear optical detection [19], the carrier-envelope phase (CEP) of the modulation waveform does not represent the CEP of the perturbation pulse, but rather the relative CEP between the fundamental and perturbation. Measuring the full electric field waveform – including the CEP – of the perturbation pulse therefore requires that the CEP of the fundamental pulse be set to zero. This, in turn, requires the use of few-cycle pulses with a duration of approximately 2.5 optical cycles or below and with stable CEP. To achieve this, we temporally compress CEP-stable mid-infrared pulses to a duration of 2.1 optical cycles using nonlinear propagation in bulk YAG and silicon windows [20-22]. Details of the nonlinear compression setup and the CEP stability of the mid-infrared pulses are described in the Methods and Supplementary Information, respectively. In this case, the modulated signal corresponds to a pulse with a duration of 24.0 fs, approximately 2.1 optical cycles, as shown in Fig. 2a. Due to the low intensity of the compressed pulses, each experimental image is integrated over approximately 100 laser shots in order to use the full dynamic range of the camera and thus increase the signal-to-noise ratio (see Supplementary Information). However, this is not an inherent limitation of the technique, and single-shot measurements of slightly longer few-cycle pulses are presented in the Supplementary Information. We confirm the validity of these measurements both by comparing to an independent measurement of the spectrum (Fig. 2b) and by measuring the spectral phase associated with the dispersion of a 2 mm-thick $CaF_2$ window placed in the perturbation pulse beam path (Fig. 2c), which agrees well with the phase calculated from Sellmeier equations [23].

The CEP of the fundamental pulse is set to zero by replacing the perturbation pulse with a weak second harmonic field, as previously described in Ref. [11]. For a sufficiently short pulse with a cosine-like waveform, multiphoton excitation is temporally confined to a single half-cycle at the center of the pulse. In this case, the detector signal will exhibit modulations at the second harmonic frequency $2\omega$. However, for sine-like pulses, or for multi-cycle pulses, the excitation events in adjacent half-cycles will be enhanced and suppressed, respectively, by the addition of the second harmonic pulse. As a result, the modulation at the second harmonic frequency is strongly suppressed, leaving instead only a weak oscillation at the fourth harmonic ($4\omega$) of the fundamental field. Figure 3 shows the modulated signals associated with a second harmonic perturbation pulse using cosine- and sine-like fundamental pulses, as well as the CEP dependent amplitude of the second harmonic modulation signal. The CEP was controlled by varying the thickness of a $CaF_2$ wedge pair placed in the incident beam. As shown, the CEP of the fundamental pulse can be set to zero simply by maximizing the amplitude of the second harmonic modulation.

We next demonstrate the single-shot measurement of laser electric field waveforms with different CEP values. We first set the CEP of the fundamental to zero, as described above, and then scan the CEP of the perturbation pulse by varying the thickness of a wedge pair placed in the perturbation pulse beam path. The results are shown in Fig. 4. By varying the CEP of the perturbation pulse by approximately π radians, we observe that the signal modulations transform from a positive to a negative sine-like waveform. The observed CEP change agrees well with that expected from the insertion of the glass wedge.

The ability to resolve the electric field waveform of a few-cycle light pulse in a single shot presents numerous opportunities to resolve attosecond dynamics in light-matter interactions [24], as well as the impulsive responses of molecules to intense ultrashort fields and their time-domain signatures [25]. Moreover, the inherent spatial resolution associated with the use of a two-dimensional detector (see Supplementary Information) is likely to enable new perspectives into the rich spatio-temporal behavior found throughout nonlinear optics. Broader applications, however, will require extension of this technique to both shorter and longer wavelengths. While the technique described here remains valid in both the multiphoton and tunneling excitation regimes, and therefore the experimental setup based on the use of a silicon image sensor is likely suitable for the detection of longer wavelength pulses, different detector technology will be required for extension to the near-infrared and potentially the visible spectrum. We have previously shown that multiphoton excitation in ZnO is suitable for wavelengths down to approximately 900 nm [14], and therefore it may be possible to spatially resolve the band fluorescence from ZnO or other dielectric materials with larger band gaps. Alternatively, AlGaN image sensors, developed for solar-blind detection in the UV [26], may provide a purely opto-electronic solution.

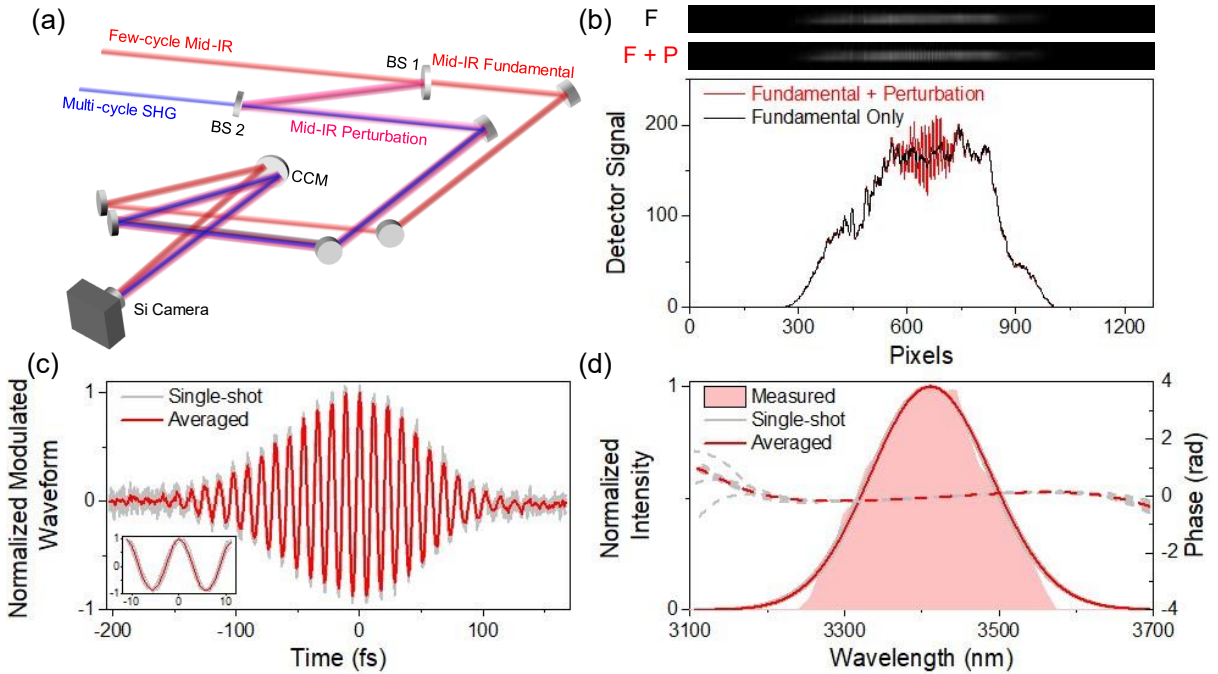

**Fig. 1 | Experimental setup and principle of single-shot measurement. a** Schematic of the experimental setup. BS: beam splitter; Mid-IR: mid-infrared; CCM: concave cylindrical mirror. The incoming laser pulse is separated into an intense fundamental pulse $E_F$ and weak perturbation pulse $E_P$, which are focused with a cylindrical mirror onto a silicon-based CMOS image sensor. For carrier-envelope phase determination, the perturbation pulse can be replaced with a second harmonic pulse. **b** Experimentally-measured single-shot image (top) and lineouts (bottom) showing the modulated nonlinear photocurrent induced by the perturbation pulse. F: fundamental; P: perturbation. **c** Measured modulation waveform after subtraction of the fundamental-only signal and normalization. **d** Spectra and spectral phases of the measured mid-infrared waveform. The spectrum measured using a grating spectrometer with a PbSe detector is shown with shaded area for comparison. The gray lines in **c** and **d** correspond to ten consecutive single-shot images obtained under identical conditions and without averaging, while the red line shows an averaged waveform, obtained by averaging over the ten images. The details of the experimental setup and normalization procedure can be found in the Methods.

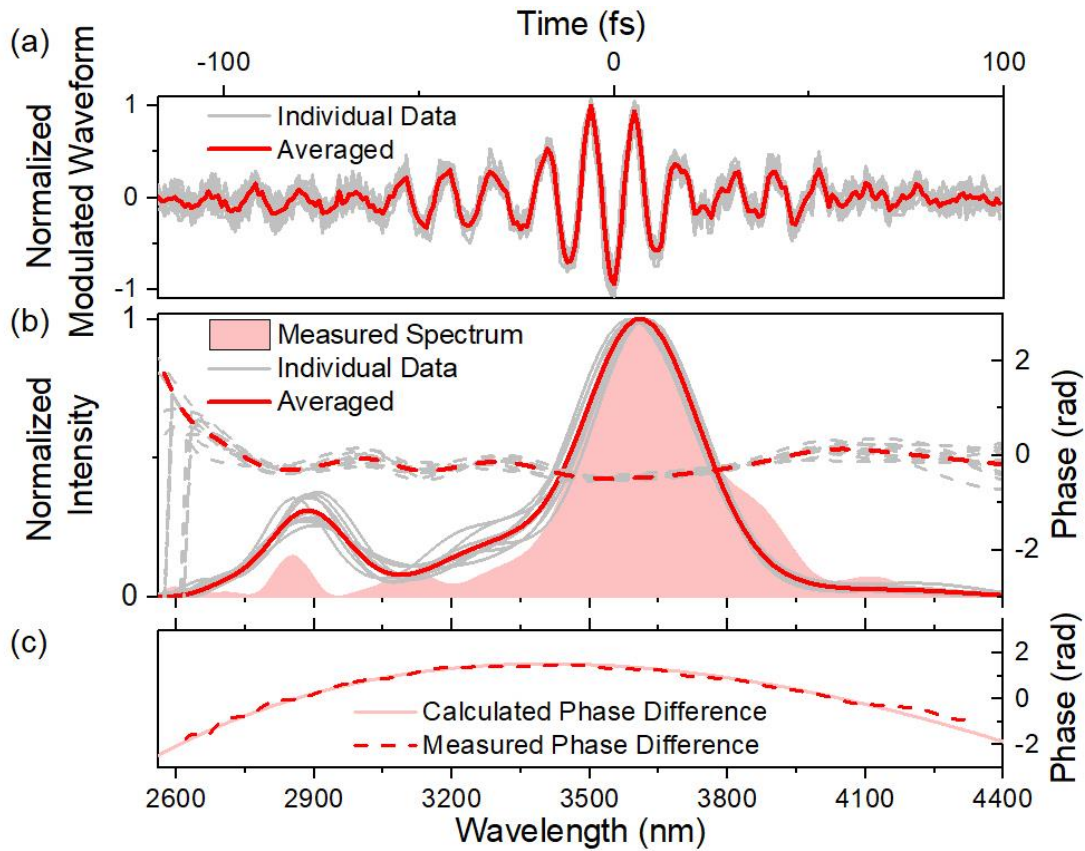

**Fig. 2 | Single-image characterization of a few-cycle pulse. a** Modulation waveform corresponding to a 2.1-cycle mid-infrared pulse generated via nonlinear compression. **b** Spectra and spectral phases of the few-cycle pulse. The spectrum measured using a home-built Fourier-transform spectrometer with a HgCdTe photodiode detector is shown with shaded area for comparison. The gray lines in **a** and **b** correspond to ten consecutive single-image measurements, each integrated over approximately 100 pulses, while the red line shows the average of the ten consecutive measurements made under identical conditions. **c** Measured spectral phase change induced by adding a 2 mm thick CaF$_2$ window in the path of the signal pulse. The retrieved phase agrees well with that predicted by Sellmeier equations [23].

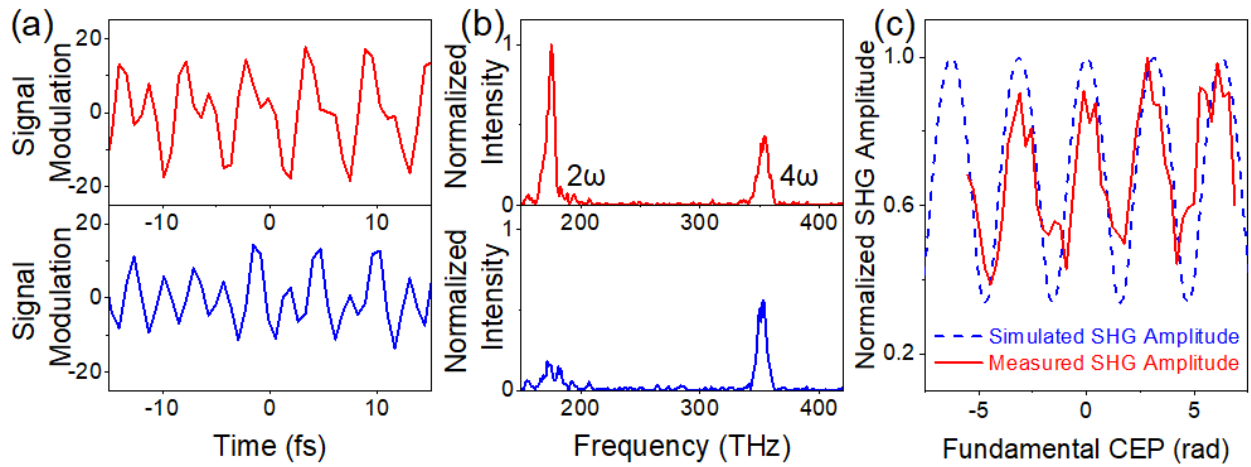

**Fig. 3 | Determination of carrier-envelope phase. a** Modulation waveforms obtained using 2.1-cycle mid-infrared fundamental pulses with different CEP and second harmonic perturbing pulses. When the fundamental pulse has a cosine-like waveform (CEP = 0, top), the modulated waveform exhibits a strong oscillation at the second harmonic frequency. However, when the fundamental pulse has a sine-like waveform (CEP = π/2 radians, bottom), the second harmonic oscillations are strongly suppressed. **b** Fourier transform of the modulations in panel a, showing oscillations at both the second harmonic (2ω) and fourth harmonic (4ω) frequencies. **c** Amplitude of the 2ω oscillation with varying fundamental CEP. The strength of the second harmonic oscillation shows a strong dependence on the fundamental CEP with a periodicity of π radians, in good agreement with simulations.

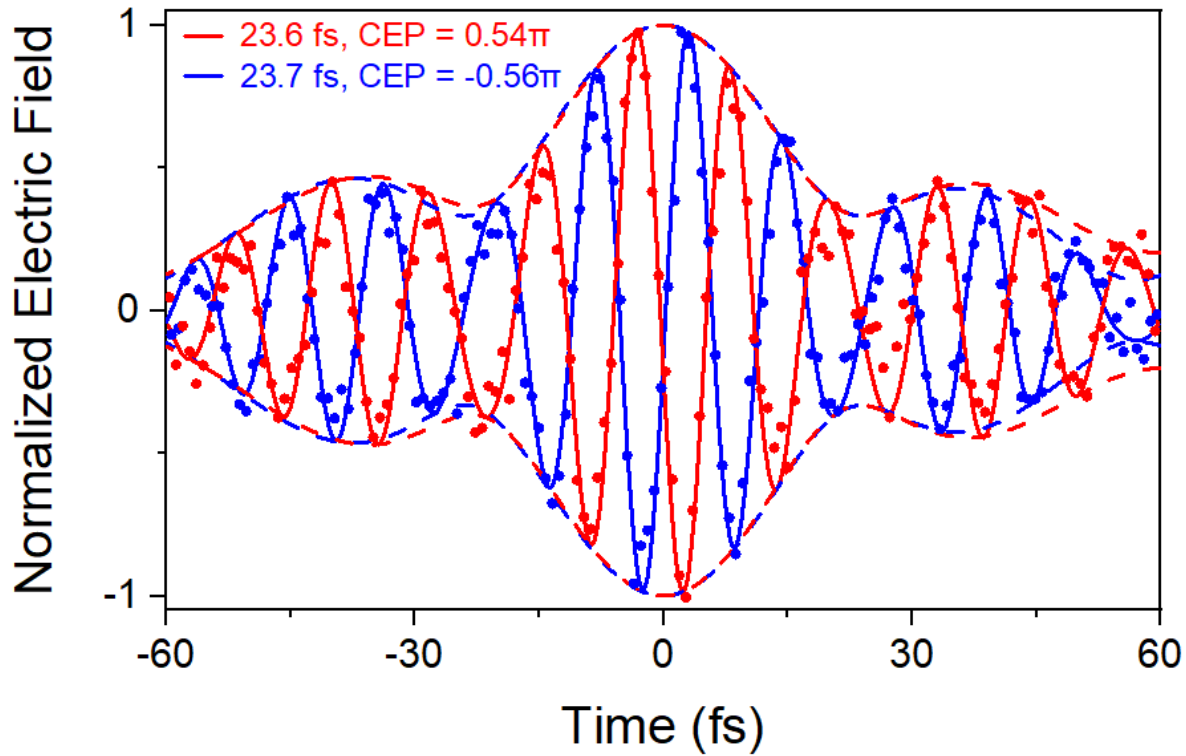

**Fig. 4 | Complete characterization of laser waveforms.** Modulation waveforms obtained using 2.1-cycle mid-infrared pulses for both the fundamental and perturbation pulses. The fundamental CEP has been set to zero, while the perturbation pulse CEP was varied by changing the insertion of a CaF$_2$ wedge. Under these conditions, the modulation waveform reflects the full electric field waveform of the perturbation pulse. Perturbation pulse CEPs of approximately ±π/2 radians are shown, which are in excellent agreement with the wedge insertion. As shown, the pulse envelope remains largely unchanged when the CEP is varied. As in Fig. 2, the waveform is obtained by averaging ten consecutive images, each integrated over approximately 100 pulses.

**Methods**

**Experimental Setup.** A schematic of the experimental setup for single-shot measurement of optical waveforms is shown in Supplementary Fig. S1. Mid-IR pulses are produced by a two-stage optical parametric amplifier (OPA, Light Conversion ORPHEUS-ONE), which is pumped by a commercial Yb:KGW amplifier (Light Conversion Carbide). The OPA idler is tunable from 2.2 – 4.5 µm, and provides approximately 10 µJ, 90 fs pulses at 10 kHz repetition rate for the wavelengths used in the experiments.

First, the single-shot on-chip detection technique was demonstrated by characterizing multi-cycle mid-IR pulses, as in Ref. [14]. To do so, OPA idler pulses with central wavelengths of 3.1, 3.4 and 3.8 µm were selected. The mid-IR pulse characterization and dispersion measurements are displayed in Supplementary Figs. S9 and S10, respectively. While the multi-cycle pulse measurement is not sensitive to the CEP of the perturbation pulse, the results confirm the good performance of the technique for pulse characterization.

Measuring the full optical waveform requires that the CEP of the fundamental pulse is set to zero. To do so requires the use of CEP-stable fundamental pulses with pulse duration below 2.5 optical cycles, and a second harmonic (SH) perturbation pulse to detect the fundamental CEP [11]. Using a $2f$-to-$f$ interferometer [27], we find that the CEP of the idler pulses is passively stable, as shown in Supplementary Fig. S7, which was expected based on the OPA geometry. For measurements with few-cycle pulses, the OPA idler wavelength was set to 3.5 µm, and the second harmonic (SH) pulses centered at 1.75 µm were generated with approximately 2% efficiency using an AlGaS2 (AGS) crystal (thickness = 1 mm) cut for type I phase matching. After the AGS crystal, a pair of CaF2 wedges was used to control the CEP of the fundamental pulse, and a dichroic mirror was used to split the mid-IR and SH pulses. For pulse compression, the mid-IR pulses were focused by a silicon lens (f = 100 mm) through three windows (2 mm thick silicon, 5 mm thick YAG, and 2 mm silicon) [22]. The first silicon window was placed 50 mm before the focal spot, while the front surfaces of the YAG and the second silicon plates were placed 2 and 10 mm after the focal spot, respectively. After the three plates, the mid-IR pulses with energy of approximately 4 µJ were collimated by a CaF$_2$ lens.

After collimation, the few-cycle mid-IR pulses were split into an intense fundamental and weak perturbation pulse using two pairs of uncoated CaF$_2$ wedges as beam splitters. The transmitted and reflected pulses, which contain 85% and 0.1% of the incident pulse energy, serve as the fundamental and perturbation arms, respectively. The second wedge pair was also used to collinearly combine the mid-IR perturbation pulse with the SH pulse with a fixed time delay, as discussed below. In the mid-IR perturbation arm, an additional wedge pair was used to balance the dispersion between the fundamental and signal arms and to vary the CEP of the perturbation pulse. The fundamental and perturbation beams were finally focused by a cylindrical concave mirror (CCM, f = 50 mm for single-shot measurements and f = 100 mm for integrated measurements) onto an 8-bit CMOS image sensor (Thorlabs DCC1545M), which serves as both the nonlinear medium and the detector. The intensity of the fundamental pulse was approximately 10 GW/cm2. The two beams overlapped on the detector with a crossing angle of approximately 3.5 degrees, allowing the time delay to be mapped onto the transverse spatial coordinate of the sensor. By varying the position of the translational stage installed in the fundamental arm, the transverse position of the maximum of the modulation waveform can be varied, and in this way, the time delay can be calibrated as shown in Supplementary Fig. S6. A 1 mm thick anti-reflection coated silicon window was placed prior to the detector, both to block the visible light and to compensate the dispersion of the CaF$_2$ wedges in the setup. After passing through the silicon plate, a pulse duration of 24.0 fs was obtained, as shown in Fig. 2.

After transmitting through the dichroic mirror, the SH pulses were reflected by a periscope in order to set their polarization parallel to that of the mid-IR pulses. After the periscope, a 2 mm thick silicon window was used to chirp the SH pulses, and a variable neutral density filter was used to control the intensity. The pulse duration of the SH pulses was estimated to be 100 fs, and the intensity was approximately 0.1% that of the mid-IR fundamental. After transmitting through the second wedge pair described above, the SH pulse propagates collinearly with mid-IR perturbation arm and is focused by the CCM onto the image sensor. As the setup uses only uncoated CaF$_2$ wedges as beam splitters, it allows broadband operation for tunable and/or few-cycle pulses, and further allows independent control of the CEP of both the fundamental and perturbation pulses. The determination of the fundamental CEP is discussed in detail in Supplementary Information.

**Normalization procedure to obtain electric field waveform.** The modulated detector signal, arising from the combined influence of the fundamental pulse $E_F$ and perturbation pulse $E_P$, can be approximated (to first order in the perturbation) as $S \propto w(E_F + E_P) \approx w(E_F) + \left(\frac{dw}{dE}\right)\Big|_{E=E_F} E_P$ [11], where $w(E)$ is the excitation rate. It is then possible to extract the field of the perturbation pulse from the difference between the detector signals with and without the perturbing field, provided that the scaling factor $\frac{dw}{dE}$ is known. While in the scanning geometry, the excitation occurs only near the peak of the fundamental pulse, and the scaling factor $\left(\frac{dw}{dE}\right)\Big|_{E=E_F}$ is constant, the single-shot measurements here are sampled over a region of the spatial profile of the fundamental pulse, and therefore the scaling factor varies across the beam profile due to the nonlinearity of the excitation rate with field strength. However, the variations induced due to nonuniformity of the fundamental pulse beam profile can be easily removed using a simple normalization procedure.

For excitation in both the multiphoton and tunneling regimes, the excitation rate can be approximated as $w \propto I^q$, where $I$ is the laser intensity and $q$ is the effective multiphoton scaling parameter. Then, $\frac{dw}{dE} = \frac{w}{E}$, which is proportional to $I^{(2q-1)/2}$. As described above, the perturbing field can be obtained from $E_P = \frac{w(E_F+E_P)-w(E_F)}{\left(\frac{dw}{dE}\right)\Big|_{E=E_F}}$, which simplifies to $E_P = \frac{S-S_F}{(S_F)^n}$, where $n = \frac{2q-1}{2q}$ and $S_F$ is the detector signal arising from the fundamental pulse only. Since $q$ can be easily measured from the dependence of the detector signal on the fundamental laser intensity (Supplementary Fig. S4), the normalization procedure can be straightforwardly applied to obtain the perturbation field, even when the fundamental beam profile varies. The normalization procedure is demonstrated in Supplementary Fig. S2, and further details of the derivation of the normalization factor are described in the Supplementary Information.

The normalization procedure described above requires successive measurements to be made with and without the perturbation pulse, and therefore imposes limits on the shot-to-shot stability of the laser pulse. However, this requirement can be relaxed, since in general the experimental geometry can be chosen such that the fundamental beam profile varies on spatial scales that are larger than those associated with the perturbation field oscillations. In this case, the measurement of $S_F$ can be replaced by a Fourier spectral filter of the perturbed signal $S$ to isolate the fast oscillations associated with the perturbation pulse from the slowly-varying background associated with the fundamental pulse.

**Data Availability**

The data that supports the plots within this paper and other findings of this study are available from the corresponding author upon reasonable request.

**Code Availability**

The codes that produced the modelled data within this paper and other findings of this study are available from the corresponding author upon reasonable request.

**Acknowledgements**

This material is based primarily on research supported by the Air Force Office of Scientific Research under Award No. FA9550-20-1-0284. S. G. was supported by the Army Research Office under Award No. W911NF-19-1-0211. We would like to thank Z. Chang for helpful discussions and for loaning the PbSe spectrometer used to measure the mid-infrared spectra.

**Contributions**

M. C. had the idea for the single-shot waveform measurement scheme and oversaw the research team. Y. L. led the experimental effort and performed most of the measurements and simulations. J. E. B. assisted with the measurements of the carrier-envelope phase dependence. J. N. and S. G. assisted with the construction of the experimental setup and with the data collection. All authors contributed to the data analysis and the creation of the manuscript.

**Corresponding author**

Correspondence to M. Chini (Michael.Chini@ucf.edu).

**Competing interests**

The authors declare no competing interests.

**Supplementary Information for:**

**Single-shot measurement of few-cycle optical waveforms on a chip**

Yangyang Liu[1], John E. Beetar[1], Jonathan Nesper[1], Shima Gholam-Mirzaei[1,2], and Michael Chini[1,3*]

[1]*Department of Physics, University of Central Florida, Orlando FL 32816, USA*

[2]*Joint Attosecond Science Laboratory (JASLab), National Research Council of Canada and University of Ottawa, Ottawa ON K1A0R6, Canada*

[3]*CREOL, the College of Optics and Photonics, University of Central Florida, Orlando FL 32816, USA*

*\*E-mail: Michael.Chini@ucf.edu*

**Normalization Factors**

The normalization procedure to obtain the electric field waveform was briefly described in the Methods, wherein the excitation rate was approximated as $w \propto I^q$, where $I$ is the laser intensity and $q$ is the effective multiphoton scaling parameter. In this case, it can easily be shown that $E_P = \frac{S-S_F}{(S_F)^n}$, where $n = \frac{2q-1}{2q}$ is the scaling factor for normalization. While this approach is certainly suitable for multiphoton processes, where the excitation rate is well-described by the power law scaling, the tunneling excitation rate is somewhat different, and the normalization factor should be reconsidered.

According to the Ammosov-Delone-Krainov (ADK) model, the rate of excitation via tunneling can be simplified as $w_{ADK} = C_1 * \left(\frac{2F_0}{E}\right)^{C_2} * e^{-\frac{2F_0}{3E}}$ [1, 2], where $C_1$, $C_2$, and $F_0$ are constants determined by the material properties and laser parameters, and $E$ is the laser field strength. Then, $\frac{dw_{ADK}}{dE} = C_3 \frac{w_{ADK}}{E} + C_4 \frac{w_{ADK}}{I}$, where $I$ is the laser intensity, and $C_3$, $C_4$ are positive constant parameters. For $C_4 = 0$, this result is the same as for the multiphoton process, and the perturbation field can be obtained from the modulation signal in the same way. On the other hand, if $C_3 = 0$, the scaling factor for normalization is found to be $n = \frac{q-1}{q}$. Therefore, in the tunneling regime, the electric field waveform can be obtained from the modulated detector signal as $E_P = \frac{S-S_F}{(S_F)^n}$, where $\frac{q-1}{q} < n < \frac{2q-1}{2q}$. Under the experimental conditions used in this paper, the multiphoton process is expected to be the dominant mechanism for excitation, and therefore the scaling factor $n = \frac{2q-1}{2q}$ was used throughout the main text. Here, however, we further investigate the influence of the normalization factor. In Fig. S3, we compare the retrieved spectra and spectral phases associated with three normalization factors: $n = \frac{q-1}{q}$, $n = \frac{2q-1}{2q}$, and $n = 1$. We find that the retrieved spectra and phases are almost the same, and the retrieved spectra all match well with the measured spectrum. This indicates that the normalized electric field waveform is not sensitive to the choice of normalization factor.

**Signal-to-Noise Ratio and Error Analysis**

In this section, we address the precision and accuracy of the measured waveform. Comparisons between multiple single-shot measurements (integration time shorter than the pulse spacing) taken under identical experimental conditions to one another, and to an integrated measurement consisting of multiple pulses (integration time longer than the pulse spacing), are shown in Figs. 1 and 2 of the main text as well as Figs. S9, S11, and S12. As shown in those figures, the agreement between each of the individual measurements is excellent, as is the agreement between the measured spectrum and spectral phase. From the repeated measurements, we can derive error bars on parameters such as the pulse duration and the relative carrier-envelope phase, which are provided in Table S1 for both the single-shot multi-cycle pulse measurements in Fig. 1 and the single-shot few-cycle pulse measurements in Fig. S11. In practice, the precision will depend also on the stability of the laser source to be measured, but this is true of any pulse or waveform measurement technique. Such instabilities can in some cases preclude the use of scanning measurements, whereas the single-shot measurement can allow the instability in the pulse parameters to be characterized. For the data presented here, the high degree of repeatability of the measurements indicates both the reliability of the technique and the high degree of stability of the laser source.

The accuracy of the waveform is influenced by the signal-to-noise ratio of the measurement, which depends upon the peak intensity of the fundamental pulse (which influences the maximum signal level), the bit-depth of the image sensor, and statistical noise in the image sensor. Here, we address the influence of each of these. The signal level recorded by the image sensor scales nonlinearly with the fundamental pulse intensity, as described in the main text and shown in Fig. S4. Therefore, through small adjustments of the fundamental intensity (by changing, for example, the focal length of the concave mirror, or the pulse energy of the incoming beam), a sufficiently high signal level can be obtained for single-shot measurements, and the signal level can be finely tuned to best take advantage of the camera bit depth. For the detector used in these measurements and a central wavelength of 3.4 µm, the tunneling current signal saturates for intensities above approximately 10-20 GW/cm$^2$. In this case, the amplitude of the modulation will then be primarily limited by the bit depth of the camera. In our experiments, we intentionally use a low-intensity perturbing pulse, as is required for the perturbation theory analysis used in the original demonstration of TIPTOE [3]. With a perturbing pulse that is approximately 1000 times weaker than the fundamental, we find that the modulation depth is approximately 15-20% of the fundamental-only signal. Therefore, as the detector used has 8 bits of dynamic range, statistical noise effectively limits the signal-to-noise ratio in our measurements to approximately 20:1, in reasonably good agreement with our measurements shown in Fig. 1 of the main text. However, this could likely be improved by using an image sensor with higher bit depth, or by using a more intense perturbing pulse together with a pulse retrieval algorithm [4]. In Fig. S11, we investigate the effects of statistical noise, by comparing single-shot measurements with averages of 4 and 9 such measurements. We find that the reduction in overall noise level scales approximately with the square root of the number of measurements included in the average, consistent with Poisson statistics.

**Table S1 | Error on pulse parameters extracted from measured waveforms.** The pulse parameter and error bar are determined as the mean and standard deviation of ten consecutive measurements.

|  | Long Pulse (Fig. 1) | | Short Pulse (Fig. S11) | |
|---|---|---|---|---|
| Amplitude (arb. units) | 1.0 | ±4.5% | 1.0 | ±6.5% |
| Pulse duration (fs) | 95.6 | ±1.1 | 28.9 | ±0.8 |
| Relative CEP (rad) | -2.19 | ±0.19 | -2.86 | ±0.17 |

Besides statistical noise, the main contributors to noise in our measurements are additive noise sources, such as those associated with dark current and readout noise. However, due to the fact that the images associated with modulated waveforms are each referenced to, and normalized, by a fundamental-only image, we do not expect additive noise in the images to be a significant contributor to the noise level of the modulated waveforms. To illustrate the effects of additive noise sources, we show in Fig. S12 a comparison of modulated waveforms obtained from an average of 9 individual single-shot measurements of a 2.3-cycle pulse, and a single-image measurement integrated over 9 consecutive pulses. In both cases, the total signal level is the same. However, for the averaged measurement, the use of multiple images increases the contribution of additive noise. We find that the single-image measurement shows a marginal improvement in signal-to-noise ratio over the averaged single-shot measurements. Together, the results of Figs. S11 and S12 suggest that the main source of noise in the measurements is due to counting statistics.

**Experimental Design Constraints**

The mapping of position to delay implicitly assumes that the wavefronts of the two beams are flat at the position of the detector. While this is a reasonable assumption, since the detector is placed at the focal distance from the curved mirror, the fact that the beams are crossed at an angle implies that it is not strictly true. In this case, geometric phase shifts in the beams as they pass through the focus may manifest themselves as gradual phase shifts across the pulse, which may be imprinted on the measured waveform. Below, we will show that these phase shifts are negligible in our measurements and discuss experimental design considerations that should be considered for measurements in other parameter spaces.

In our measurements, the camera is put at the focus where the wavefront is approximately flat. The delay range over which the modulations are observed varies from approximately 200 fs for the few-cycle pulses to 600 fs for the multi-cycle pulses. These values correspond to maximum longitudinal distance between the two foci of 0.07 and 0.2 mm, respectively. Under the assumptions of gaussian focusing with the measured beam radius of $w_0$ = 30 µm for a cylindrical mirror with 50 mm focal length, we estimate the total phase shift associated with the Gouy phase and diffraction to be less than 240 mrad for the multi-cycle pulses and 80 mrad for the few-cycle pulses. This finding is consistent with Figs. S2 and S6, which show the reproducible waveform measurements and a linear time delay shift across the beam profile, respectively. For the integrated waveform measurements shown in Figs. 2, 3, and 4, a cylindrical mirror with a longer focal length of $f$ = 100 mm was used. In that case, the phase shift is expected to be below 20 mrad across the waveform.

The potential contributions of geometrical phases to the waveform measurement, should, however, be considered when designing a pulse characterization system for pulses with lower energy, longer wavelength, or longer pulse duration. This is due to the fact that the geometrical phase shifts will necessarily be larger when using focusing optics with shorter focal lengths, longer wavelengths, or larger crossing angles. In all cases, the experimental conditions should be chosen such that the spatial phase variation does not significantly impact the waveform measurement.

**Determination of the fundamental CEP**

To characterize the electric waveform (including CEP) of the perturbation pulse, it is necessary to set the CEP of the fundamental pulse to zero, which can be achieved by replacing the mid-IR perturbation pulse with a second harmonic (SH) pulse [3]. The photocurrent signals generated with fundamental only, SH only, and fundamental combined with SH, are plotted in Fig. S8(a). We note that the intensity of the second harmonic pulse is approximately 0.1% of the fundamental. Though the intensity of the SH pulse is weak, the excitation rate is much higher than for the fundamental since the excitation is a two-photon, rather than four-photon, process. When the fundamental and the SH pulses are spatially, but not temporally, overlapped (noted as "Not Overlapped" in the figure legend), the total photocurrent signal is the sum of the photocurrents generated by fundamental only and SH only. However, when the fundamental and the second harmonic pulses are overlapped both spatially and temporally (noted as "Overlapped 1" and "Overlapped 2" in the figure legend), the photocurrent signal is enhanced. By subtracting the "Not Overlapped" photocurrent signal from the "Overlapped" photocurrent signals, the difference signals displayed in Fig. S8(b) can be obtained. The difference signals contain both a strong, slowly varying background signal and a weak modulation signal. After taking the Fourier transform of the difference signals, there are three main frequency components, located at zero frequency, at the second harmonic (2ω) oscillation frequency, and at the fourth harmonic (4ω) oscillation frequency. The zero-frequency component corresponds to the slowly varying background, which we attribute to the excitation associated with two photon absorption of one mid-IR photon on the short-wavelength tail of the spectrum and one second harmonic photon. The 2ω and 4ω oscillations, however, can be viewed as a weak perturbation induced by the SH pulse, which can be used to determine the fundamental CEP as demonstrated in Ref. [3].

We have simulated the influence of the SH perturbation pulse to the multiphoton excitation of the mid-IR fundamental, as shown in Fig. S8(d). When setting the fundamental CEP to zero, the modulation signal appears as a strong 2ω oscillation combined with a weak 4ω oscillation. However, if the fundamental CEP is set to ±π/2, the 2ω oscillation becomes weaker than the 4ω oscillation. By taking the Fourier transform of the modulation signals, the simulated spectra are shown in Fig. S8(e), which are in good agreement with the experiments in Fig. 3(b) in the main text. By tuning the thickness of the wedge pair installed before the dichroic mirror, the CEP of the fundamental pulse can be continuously controlled. We have simulated and measured the dependence of the 2ω oscillation amplitude on the CEP of the fundamental pulse, as shown in Fig. S8(f). The measured result agrees well with the simulation, and the CEP of the fundamental pulse can be easily set to zero by optimizing the second harmonic amplitude to its peak.

**Measurement of Spatiotemporal Coupling**

The single-shot detection scheme is based upon mapping the time axis to one of the spatial axes of the image sensor. However, because we use a two-dimensional sensor, the orthogonal axis retains information about the spatial profile of the pulse. Here, we show that this allows us to characterize spatiotemporal couplings in the pulse, and therefore to evaluate the presence of spatiotemporal distortion in our compressed pulses.

To demonstrate the sensitivity of the measurement to spatiotemporal coupling, we imprint a dynamic wavefront rotation [5] on the perturbing pulse by inserting a $CaF_2$ wedge (6° wedge angle) into the beam [6], as shown schematically in Fig. S13(a) and (d). The insertion of the wedge leads to a wavefront rotation at the focus, which results in a fan-like structure in the space-time structure of the electric field that can be resolved by measuring the dependence of the waveform along the vertical dimension of the image sensor. As shown in panels (b) and (e), the wavefront rotation is evident in our measurement, and it can be controlled by changing the direction of the wedge insertion. The Fourier transforms of the measured waveforms reveal a spatial chirp in the beam, as shown in panels (c) and (f). The degree of spatial chirp that we observe in the Fourier transformed waveforms is in reasonably good agreement with an estimate of the spatial chirp estimated from a ray tracing analysis under the experimental conditions.

Lastly, we perform a similar analysis on few-cycle pulses generated via nonlinear spectral broadening as in Fig. 2. Fig. S14 shows the spatially-resolved waveform for few-cycle perturbation pulses without added wavefront rotation (i.e., with the wedge removed). The result shows a negligible amount of spatiotemporal distortion, and an analysis of the spatially-resolved spectrum indicates that the on-axis components of the compressed beam are spectrally homogeneous.

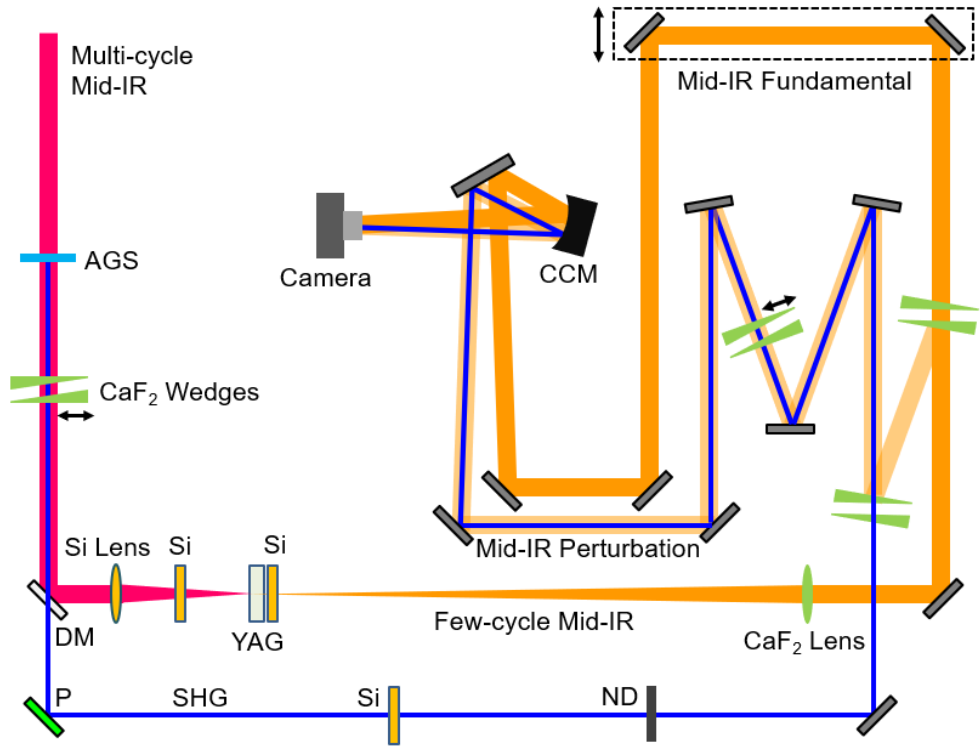

**Fig. S1 | Experimental setup for single-shot measurement of optical waveforms.** The experimental setup is described in detail in the Methods. Mid-IR: mid-infrared; AGS: AlGaS$_2$ second harmonic generator; DM: dichroic mirror; P: periscope; SHG: second harmonic beam; Si: silicon; ND: variable neutral density filter; CCM: concave cylindrical mirror.

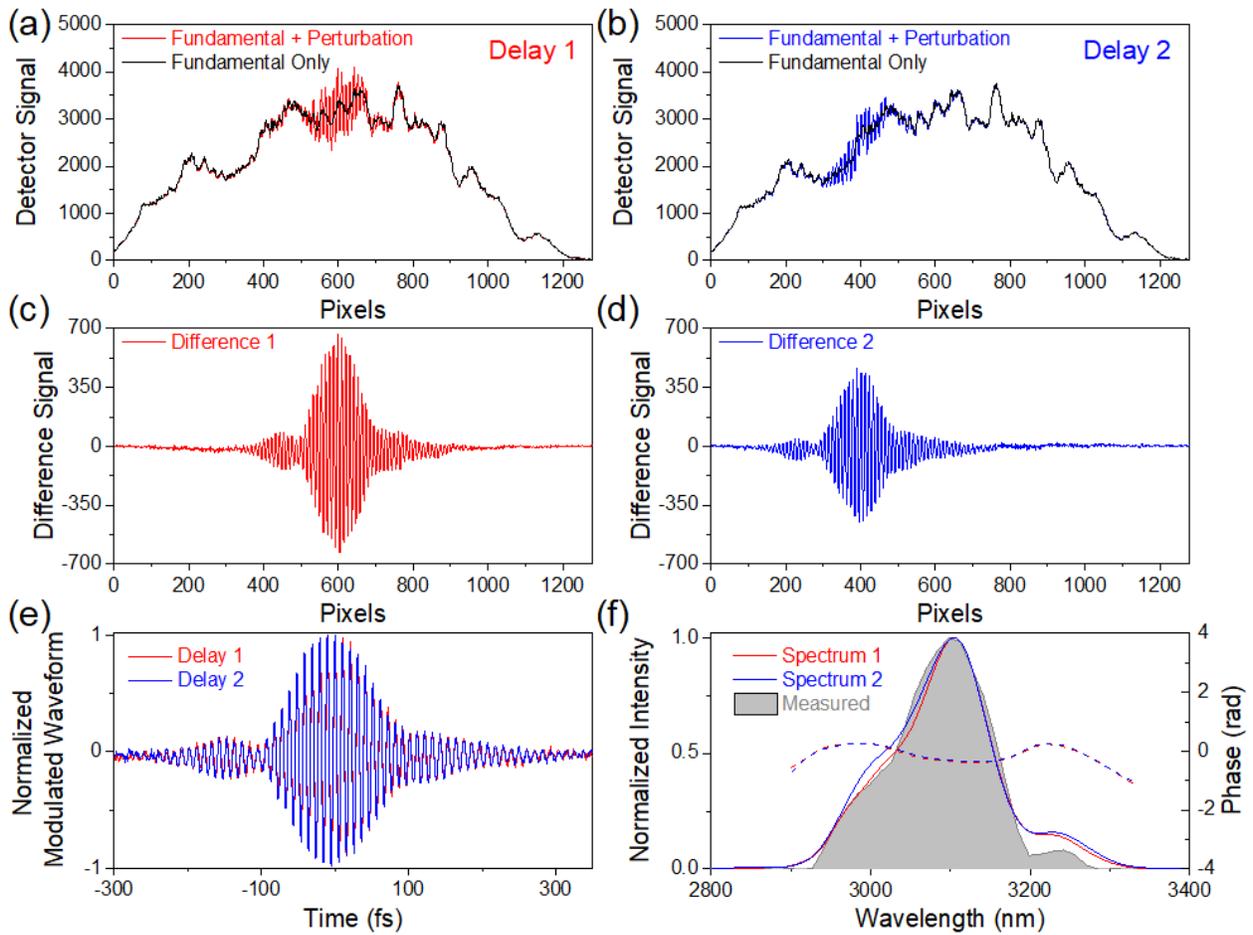

**Fig. S2 | Normalization procedure to obtain electric field waveform. a**, **b** Measured detector signals obtained with fundamental only (black line) and fundamental plus perturbation (red and blue lines, respectively) at different delay positions. **c**, **d** Difference signals, which are obtained by subtracting the fundamental only signal from the measured modulated signal, at different delay positions. Note that there are significant differences between the difference waveforms. **e** Normalized modulated waveforms at different delay positions. Note the good agreement between the normalized waveforms. **f** Retrieved spectra (solid lines) and phases (dashed lines) at different delay positions, measured spectrum (shaded area).

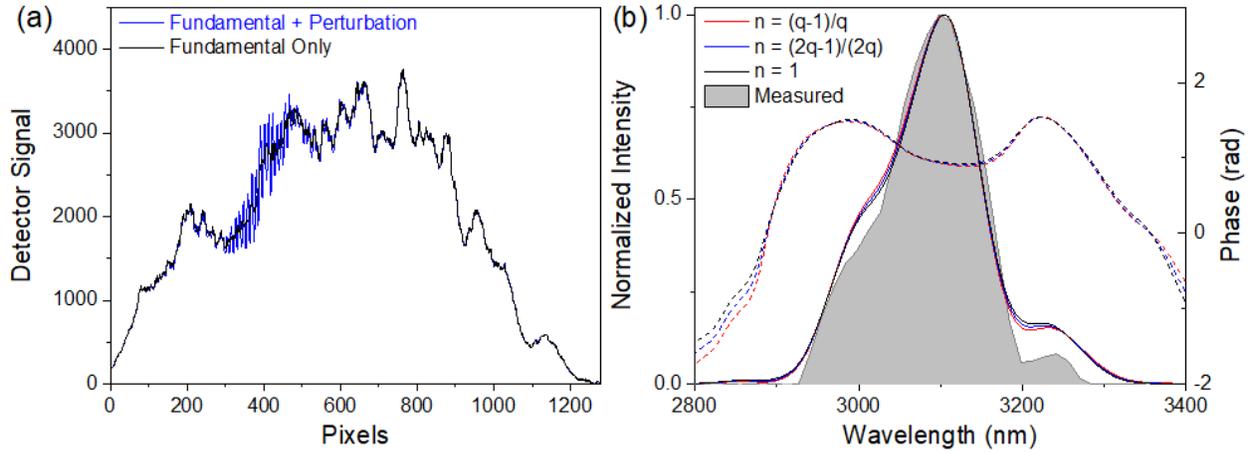

**Fig. S3 | Influence of the normalization scaling factor. a** Detector signals generated by fundamental only (black line) and fundamental plus perturbation (blue line). **b** Retrieved spectra (solid lines), retrieved phases (dashed lines), and measured spectrum (shaded area) obtained with three different normalization factors. To compare the performance of different normalization factors under adverse conditions, a steeply varying region of the beam profile was chosen intentionally. The retrieved spectra and phases are almost the same for the three normalization factors.

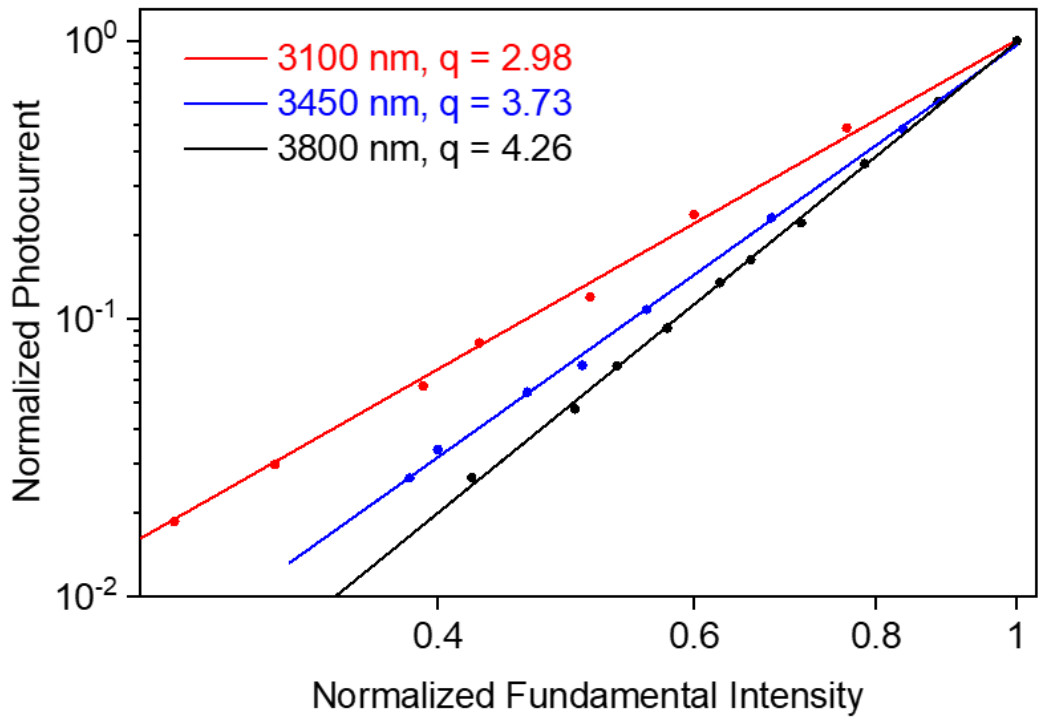

**Fig. S4 | Dependence of the detector signal on fundamental intensity.** Normalized data are shown as dots, with power-law fits shown as lines. As described in the main text, the excitation rate can be expressed as $w \propto I^q$, where $I$ is the laser intensity and $q$ is the effective multiphoton scaling parameter. For central wavelength of 3.1, 3.45 and 3.8 μm, q is calculated to be 2.98, 3.73 and 4.26 based on the fitting, respectively.

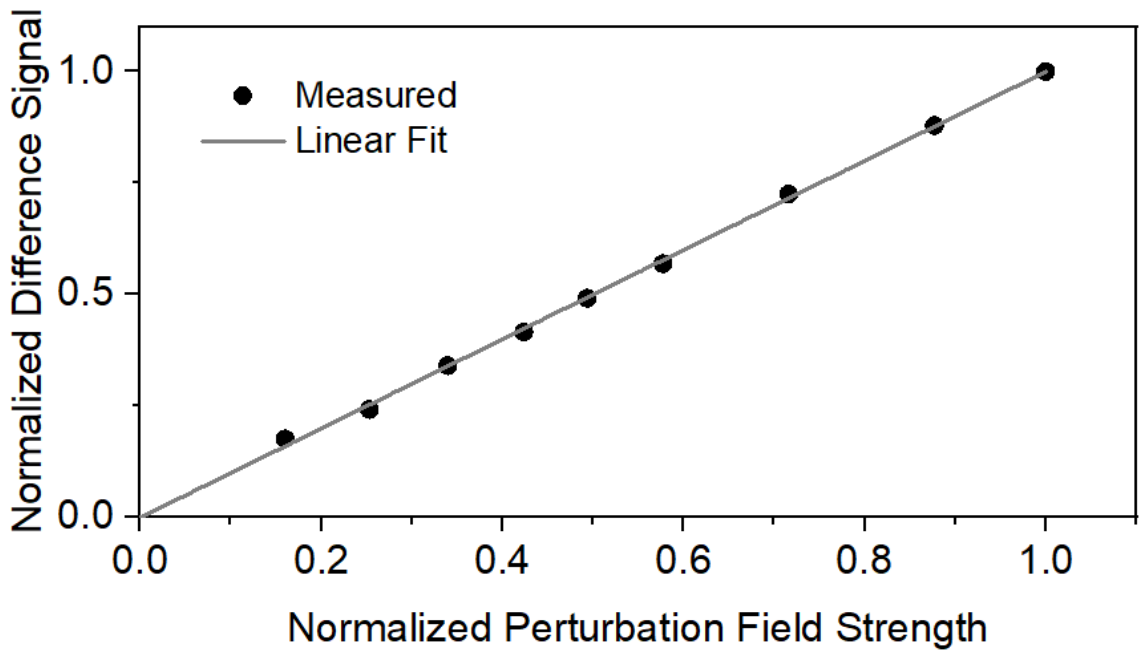

**Fig. S5 | Dependence of the difference signal on perturbation field strength.** The perturbation field strength is controlled by adding a rotational neutral density filter in the perturbation arm. For a particular position on the sensor (i.e., a particular value of fundamental-signal time delay), the difference signal, which is obtained by subtracting the fundamental only signal from the modulated signal, scales linearly with the electric field amplitude.

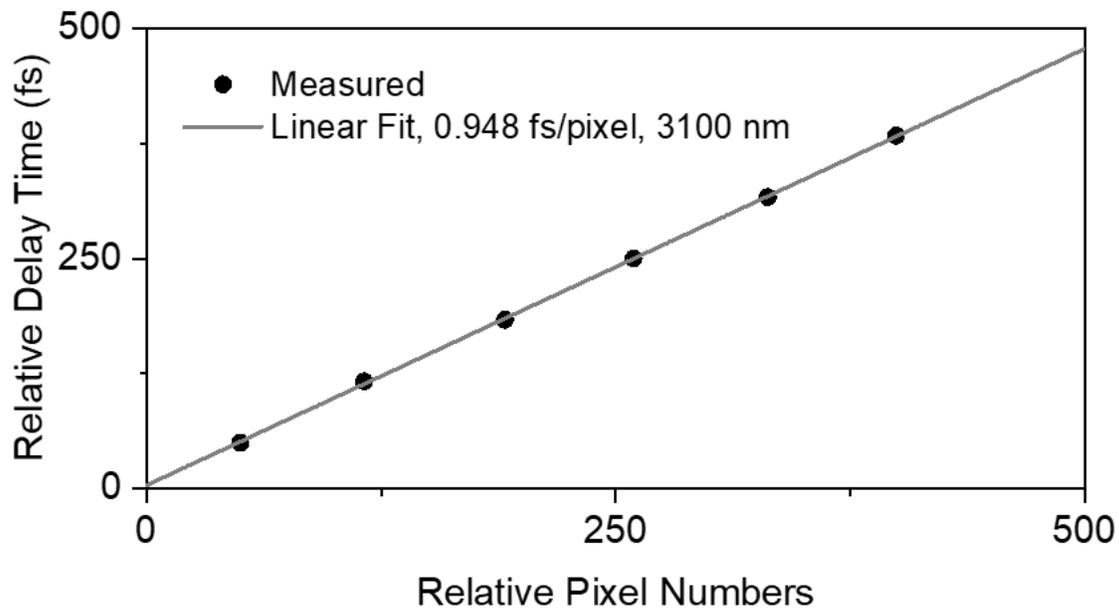

**Fig. S6 | Time delay calibration.** Measured data are shown as dots with a linear fit. By varying the delay between the fundamental and perturbation pulses, the maximum of the normalized modulated waveform will appear at different detector positions. As shown here for the multi-cycle 3.1 μm driving pulses, the time delay step was calibrated to be 0.948 fs/pixel.

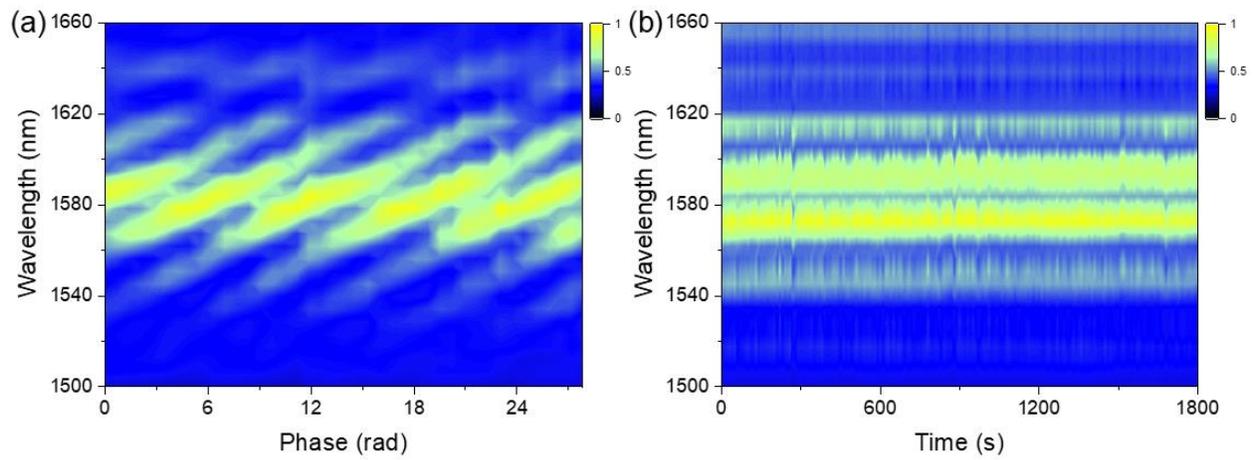

**Fig. S7 | CEP stability of the OPA idler. a** Dependence of the 2f-f interferogram on the CEP of the OPA idler centered at 3.2 μm. The CEP of the OPA idler was changed by varying the thickness of the wedge pair, which is installed in front of the 2f-f interferometer. **b** Interference fringes measured in the 2f-to-f interferometer over 30 minutes. The interference fringes indicate that the CEP of the OPA idler is stable.

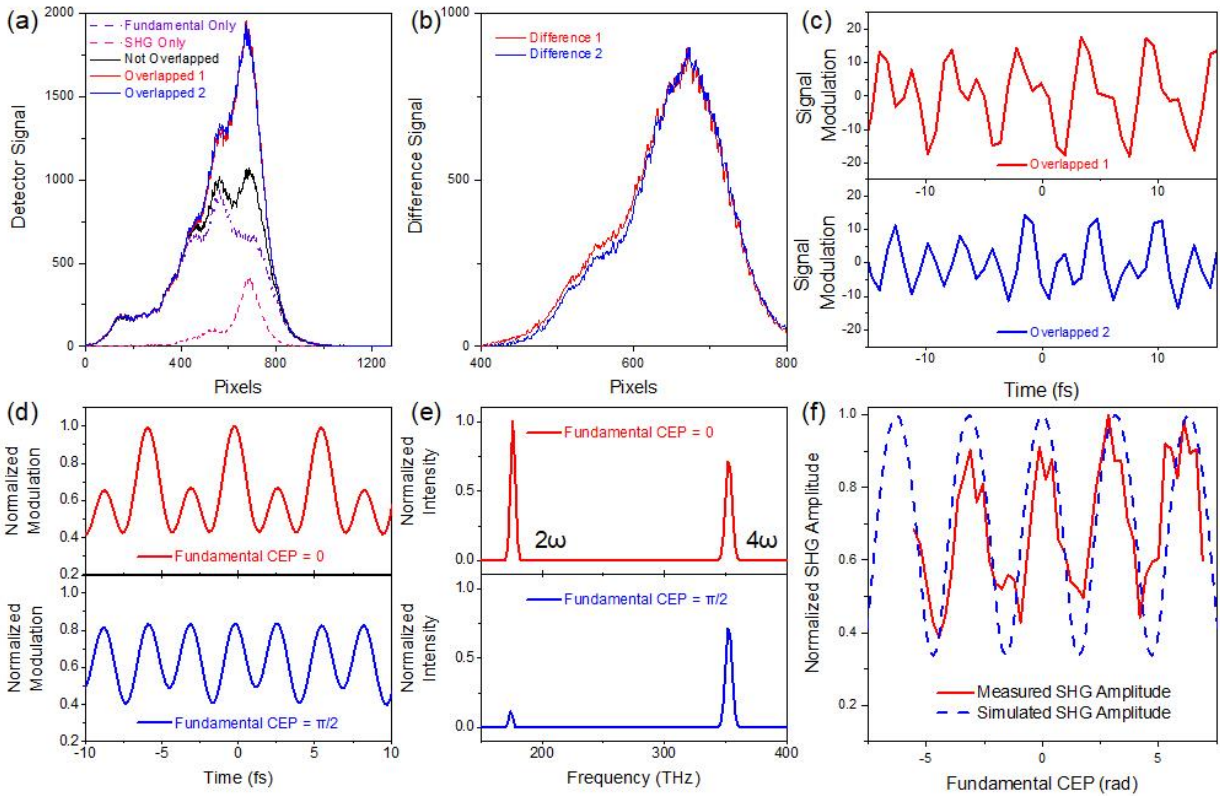

**Fig. S8 | Determining the CEP of the fundamental pulse. a** Detector signals generated by fundamental only (dashed purple line), second harmonic only (dashed pink line), and fundamental plus second harmonic pulses (solid lines). **b** Difference signals obtained by subtracting the signal without temporal overlap from the temporally overlapped signal. The signal includes both a strong slowly varying background and with a weak modulation at the second and fourth harmonic frequencies. **c** Reconstructed modulations by considering a spectral window spanning the second and fourth harmonic frequencies. **d** Simulated modulations when the CEP of fundamental is set to zero (red line) and π/2 (blue line). **e** Retrieved spectra obtained by taking Fourier transform of the simulated normalized modulations. **f** The measured (solid red line) and simulated (dashed blue line) second harmonic amplitude.

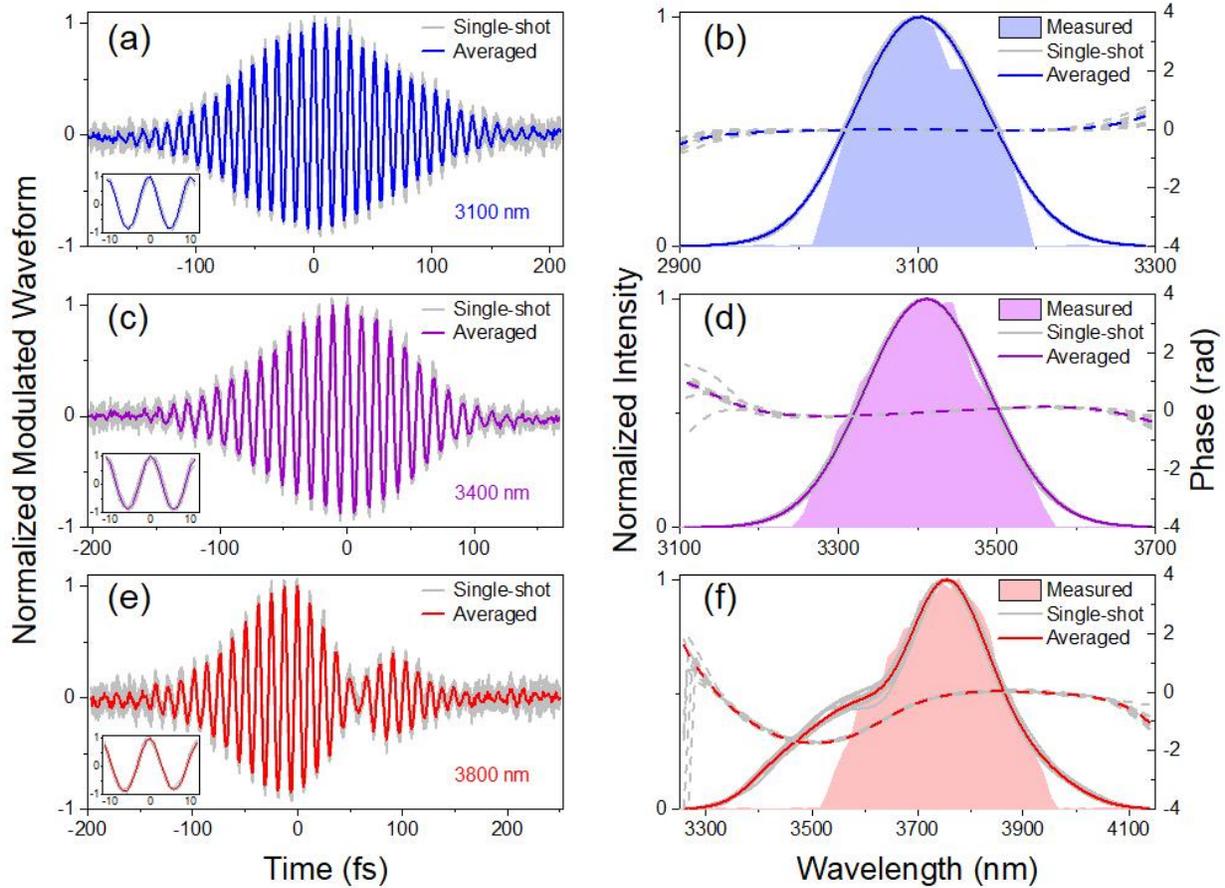

**Fig. S9 | Single-shot measurement of multi-cycle mid-IR pulses. a**, **c**, **e** Normalized modulated waveforms corresponding to 3.1, 3.4, and 3.8 μm central wavelengths, respectively. **b**, **d**, **f** Measured spectrum (shaded area), retrieved spectra (solid lines), and retrieved spectral phases (dashed lines). The gray lines represent ten individual single-shot measurements, while the colored lines indicate averages of the ten single-shot measurements under identical conditions.

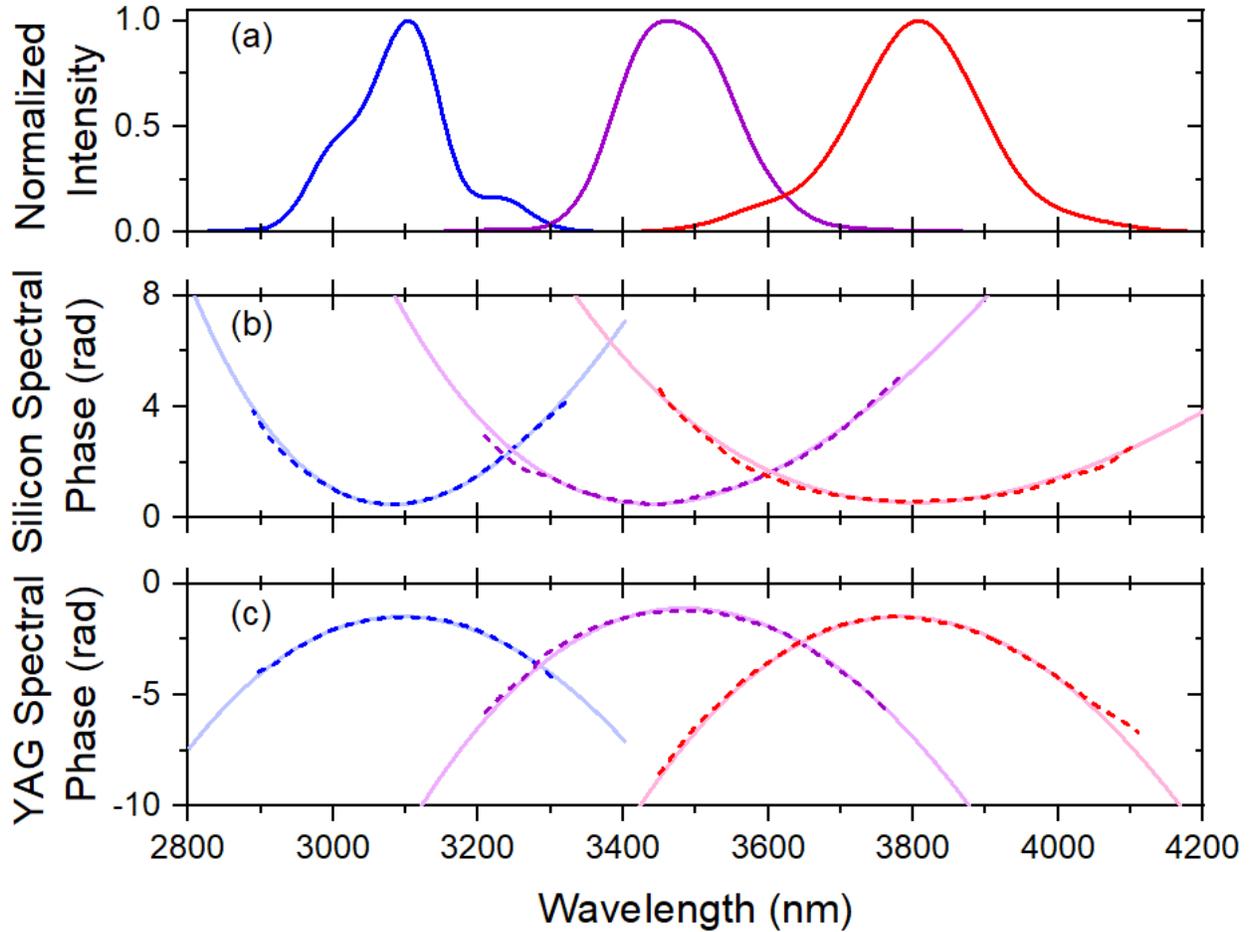

**Fig. S10 | Dispersion measurement with multi-cycle mid-IR pulses. a** Retrieved spectra obtained from the normalized modulated waveforms corresponding to 3.1, 3.45, and 3.8 μm central wavelengths, respectively. **b** Measured (dashed line) and calculated (solid line) spectral phase change induced by 8 mm Si. **c** Measured (dashed line) and calculated (solid line) spectral phase change induced by 8 mm YAG. These measurements were done by performing additional measurements in which Si or YAG windows (total thickness = 8 mm) were added in the perturbation arm, and comparing the spectral phases retrieved with and without the additional dispersion. The induced spectral phase changes (solid lines) are calculated based on the Sellmeier equations of Si and YAG [7-9].

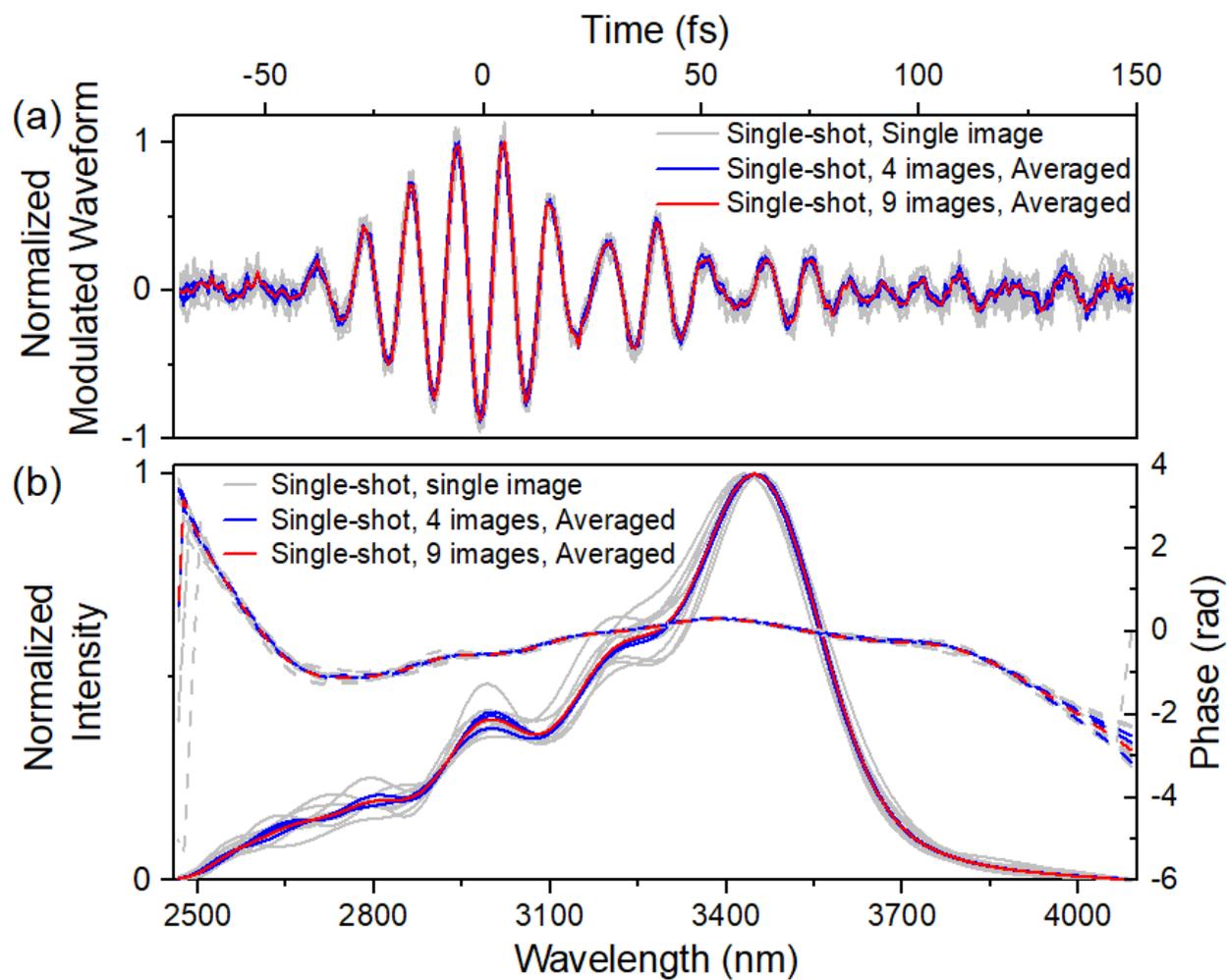

**Fig. S11 | Comparison of single-shot and averaged waveforms. a** Measured waveforms corresponding to a 2.5-cycle pulse. The comparison of individual single-shot measurements (gray, 10 lines) with multiple averaged measurements consisting of four (blue, 3 lines) or nine (red, 1 line) single-shot images shows how the signal-to-noise ratio improves with the addition of additional images. The scaling of the signal-to-noise ratio with the integrated signal level suggests that the primary source of noise is statistical in nature. **b** The retrieved spectra and phases of the single-shot and averaged measurements are in good agreement with one another.

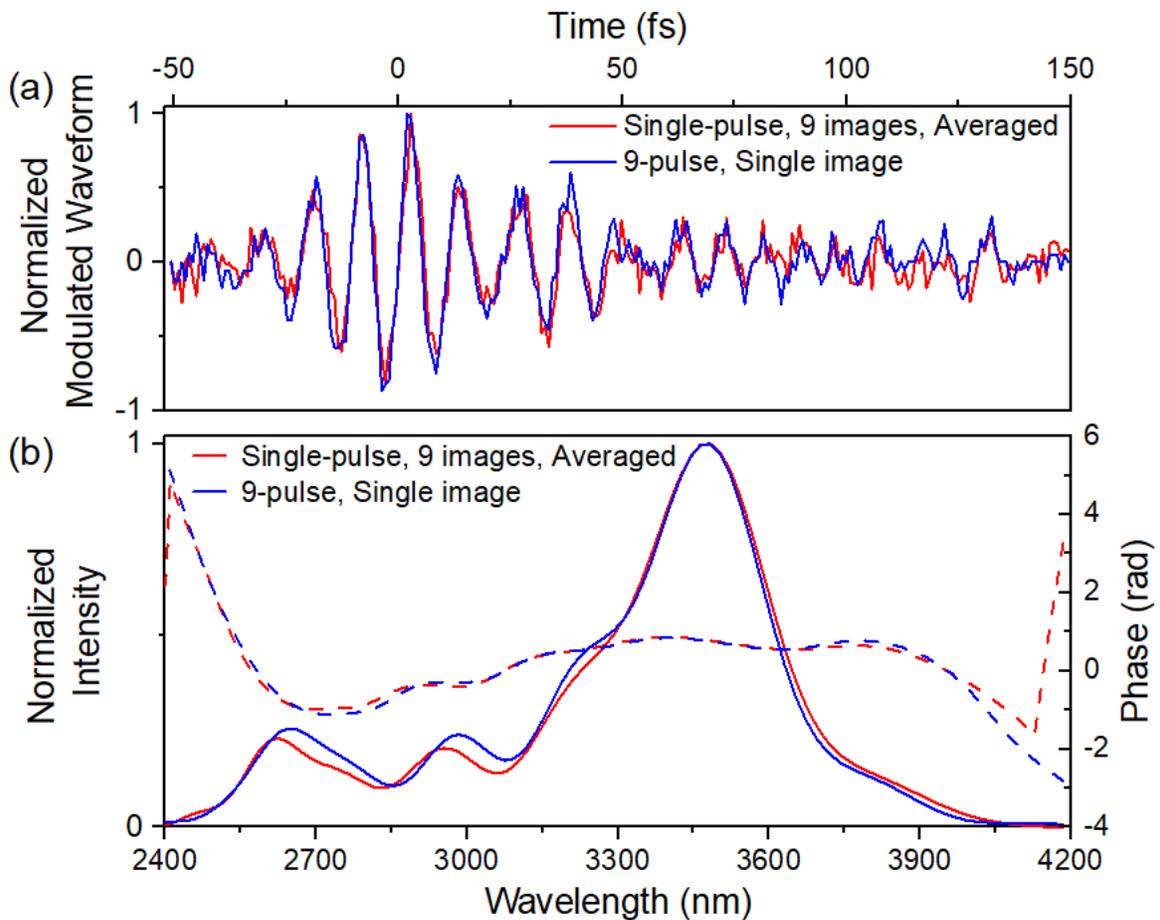

**Fig. S12 | Comparison of single-shot and multi-shot measurements. a** Measured waveforms corresponding to a 2.3-cycle pulse. In this case, we compare a single image, integrated over 9 pulses (blue) with the average of 9 single-shot images, under otherwise identical conditions. The averaged single-shot measurement has a higher noise level than the multi-shot single-image measurement, indicating the contribution of additive noise. The signal level for the single-shot images was intentionally kept low here in order to have the same statistical noise for the multi-shot and averaged single-shot measurements. **b** The retrieved spectra and phases of the single-shot and averaged measurements are in good agreement with one another.

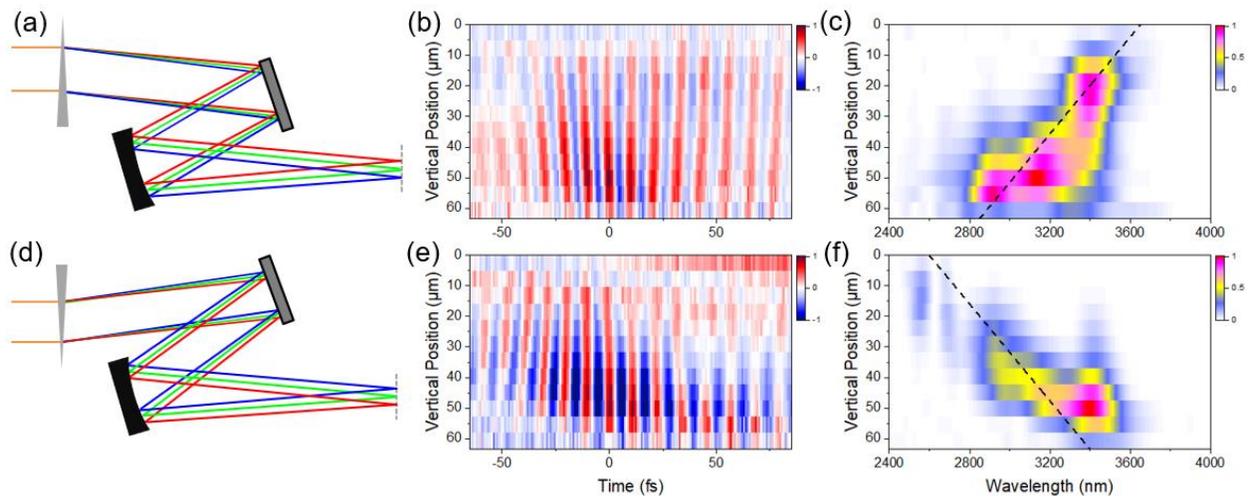

**Fig. S13 | Characterization of spatiotemporal coupling. a** Schematic setup with rays representing different wavelengths in the pulse. **b** Spatially-resolved waveform measurement exhibiting wavefront rotation. **c** Fourier transform of panel b, showing spatially-resolved spectrum along with a calculation of the spatial chirp (dashed line). **d**, **e**, **f** Same as a, b, c but with the wedge inserted in the opposite direction.

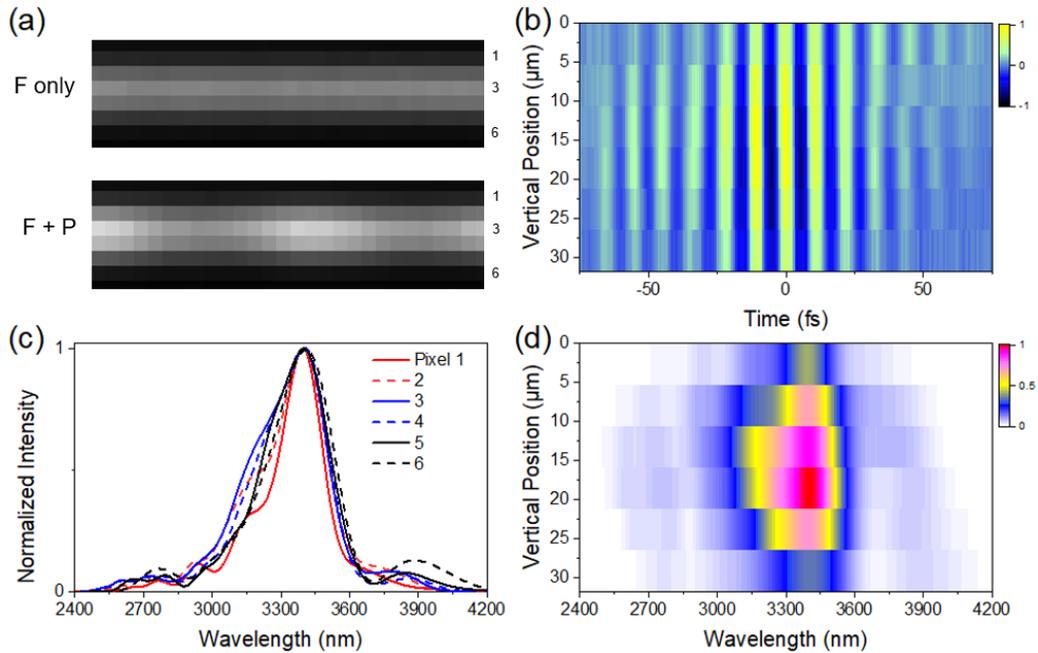

**Fig. S14 | Characterization of spatiotemporal distortions in self-compressed pulses. a** Images showing the tunneling current arising from the fundamental pulse only (top) and from the combination of the fundamental and perturbation (bottom). **b** Spatially-resolved waveform measurement obtained from the images in panel a. Unlike in Fig. S13, the modulations at different vertical positions are in phase with one another. **c** Normalized spectra obtained by taking the Fourier transform along the time axis in panel b. The good agreement between the spectra indicates a lack of spatial chirp in the beam. **d** Fourier transform of panel b, showing the variation in the spectral amplitude across the beam profile. From the measurement, we can conclude that there is negligible spatiotemporal distortion in the self-compressed beam.